\documentclass[letterpaper,twocolumn,showpacs,prl,aps]{revtex4-1} 

\usepackage{amsmath}  
\usepackage{amsfonts} 
\usepackage{graphicx} 
\usepackage{amssymb}
\usepackage{subfig}
\usepackage{float}
\usepackage{xcolor}
\usepackage{caption}
\usepackage{hyperref}

\begin{document}

\title{Vortex wake patterns in superfluid $^{4}He$}

\author{Eugene B. Kolomeisky}

\affiliation
{Department of Physics, University of Virginia, P. O. Box 400714,
Charlottesville, Virginia 22904-4714, USA\\ email \href{mailto:ek6n@virginia.edu}{ek6n@virginia.edu}}

\date{\today}

\begin{abstract}
Excitations in the form of quantized vortex rings are known to exist in superfluid $^{4}He$ at energies and momenta exceeding those of the Landau phonon-roton spectrum.  They form a vortex branch of elementary excitations spectrum which is disconnected from the Landau spectrum. Interference of vortex ring excitations determines wake patterns due to uniformly traveling sources in bulk superfluid at low speeds and pressures.   The dispersion law of these excitations resembles that of gravity waves on deep water with infrared wave number cutoff.  As a result, vortex wake patterns featuring elements of the Kelvin ship wake are predicted.  Specifically, at lowest speeds the pattern with fully developed transverse and diverging wavefronts is present.  At intermediate speeds transverse wavefronts are absent within a cone whose opening angle increases with the source velocity. At largest speeds only diverging wavefronts confined within a cone whose opening angle decreases with the source velocity are found.  When experimentally observed, these changes in appearance of wake patterns serve as indicators of the beginning part of the vortex branch of elementary excitations.             

\end{abstract}

\maketitle

Landau's insight that elementary excitations determine low-energy properties of many-body interacting systems is a paradigm of physics \cite{LL9}.  One such property probing elementary excitations spectra that is common to a number of physical systems is far-field wake patterns representing response to uniformly traveling external disturbances.  Familiar examples include Mach waves due to a supersonic projectile~\cite{LL6}, Cherenkov radiation emitted by a rapidly moving charge~\cite{LL8}, and ship waves~\cite{Kelvin, Lamb}, all of which are examples of coherent generation of the medium's elementary excitations~\cite{wake_review}. 
 
If the relevant elementary excitations are characterized by the dispersion law $\omega(\textbf{k})$ (where $\omega$ is the frequency and $\textbf{k}$ is the wave vector), a wake is present whenever there is a wave mode whose phase velocity $\textbf{k}\omega/k^{2}$ (here $k$=$|\textbf{k}|$) matches the projection of the velocity of the source $\textbf{v}$ onto the direction of radiation $\textbf{k}/k$ \cite{Lamb,wake_review}.  For a source moving with velocity $\textbf{v}$, this requires the existence of a wave vector $\mathbf{k}$ satisfying the Mach-Cherenkov-Landau (MCL) resonant radiation condition
\begin{equation}
\label{Cherenkov}
\omega(\textbf{k})=\textbf{k}\cdot\textbf{v}\equiv kv\cos\varphi.
\end{equation}     
Here $v=|\textbf{v}|$ and $\varphi$ is the angle between the vectors $\textbf{k}$ and $\textbf{v}$.  Eq.(\ref{Cherenkov}) also describes the onset of Landau damping in a plasma \cite{LL10} and the breakdown of superfluidity \cite{LL9}.

When the excitation spectrum is linear, 
\begin{equation}
\label{sound}
\omega=uk
\end{equation}
where $u$ is the speed of sound (or light), the condition (\ref{Cherenkov}) becomes $\cos \varphi=u/v$.  It can be satisfied only if $u\leqslant v$, i.e. there is a finite critical velocity to generate a wake pattern, $v_{c}=u$.  

Recently developed theory \cite{KCS} makes it possible to understand wake patterns due to excitations of general isotropic dispersion laws.   One of its applications included wake patterns in superfluid $^{4}He$ produced by a small uniformly moving source or equivalently by a stationary obstacle in the presence of background flow.  The input to the theory \cite{KCS} is the excitation spectrum of $^{4}He$.  

The latter, as was recognized by Landau \cite{Landau}, is a superposition of two continuous spectra:  one corresponding to potential flow and one, at a higher energy, corresponding to vortex motion.  The potential flow part of the spectrum known as the Landau phonon-roton spectrum has the following properties \cite{Landau,LL9}:

For small wave numbers $k$ the elementary excitations are phonons with a linear spectrum (\ref{sound}).  As the wave number increases, the function $\omega(\textbf{k})=\omega(k)$ reaches a maximum, followed by a "roton" minimum at some $k_{0}$.  In the vicinity of $k=k_{0}$ it is customary to expand the spectrum in powers of $k-k_{0}$:
\begin{equation}
\label{roton}
\omega=\frac{\Delta}{\hbar}+\frac{\hbar (k-k_{0})^{2}}{2\mu}
\end{equation}
where $\Delta$, $\mu$ and $k_{0}$ are empirically known parameters depending on the pressure \cite{pressure} such as with good accuracy the critical velocity to generate a wake pattern or equivalently to destroy superfluidity is given by the slope of the straight line connecting the origin $\omega(0)=0$ to the roton minimum $\omega(k_{0})=\Delta/\hbar$ \cite{KCS}:
\begin{equation}
\label{roton_velocity}
v_{c}=\frac{\Delta}{\hbar k_{0}}.
\end{equation}

The Landau critical roton velocity (\ref{roton_velocity}) has been attained only in experiments with isotopically pure $^{4}He$ at a pressure exceeding $12$ bars \cite{vortex}.  Likewise, roton wake patterns predicted in Ref. \cite{KCS} can be observed in their pure form only at elevated pressure.  The reason being is that excitations of vortex nature, namely vortex rings \cite{Feynman}, play dominant role  at low pressures.  As a whole vortex rings have both definite momentum and energy, and thus are a special type of elementary excitations \cite{LL9}.   

Originally Landau argued \cite{Landau} that the vortex motion branch of the elementary excitations spectrum starts out according to Eq.(\ref{roton}) with $k_{0}=0$. However experimental data on second sound indicated that it is Eq.(\ref{roton}) that describes short-wavelength part of the potential flow branch of the excitation spectrum;  the roton does not represent vortex motion as its group velocity is zero while vortex ring cannot be at rest.  A qualitative theory of the Landau phonon-roton spectrum has been given by Feynman and its improvements have been proposed since then \cite{Feynman}.

Landau also pointed out that just as there is no continuous transition in quantum mechanics between states with zero angular momentum and states with finite angular momentum, there may not be a continuous transition between the potential flow and the vortex motion branches of the excitation spectrum \cite{Landau}.  The phonon-roton spectrum is known to end at $\omega=2\Delta/\hbar$, $k=k_{c}$ \cite{LL9}.  Specifically, extrapolated to zero pressure values of the parameters are $\Delta=0.74 ~meV$, $k_{0}=1.9~\AA^{-1}$, and $k_{c}=3.6~ \AA^{-1}$ \cite{pressure}.  Pitaevskii \cite{Pitaevskii66} discussed a possibility and Marchenko and Parshin \cite{MP} further argued that at low pressure the spectrum of vortex rings begins at a wave number $k_{b}>2k_{0}$ and an energy $\hbar\omega_{b}$ of several $\Delta$.  The coordinates of the beginning point of the spectrum were estimated as $k_{b}= 4.3~\AA^{-1}$ and $\hbar\omega_{b}=2.2~meV$ \cite{MP}.

In the macroscopic approximation the spectrum of quantized vortex rings is given by \cite{LL9}
\begin{equation}
\label{vortex_ parametric}
\omega=2\pi^{2}\frac{\rho_{s}\hbar}{m^{2}}R\ln\frac{R}{a},~~~~~k=2\pi^{2}\frac{\rho_{s}}{m}R^{2}
\end{equation}
where $\rho_{s}$ is the superfluid density, $m$ is the mass of the $He$ atom, $R$ is the radius of the vortex ring playing a role of the parameter, $a$ is the core radius of the vortex having atomic scale, and it is assumed that $\ln(R/a)\gg1$.  Eqs.(\ref{vortex_ parametric}) conveys the fact that the larger the ring radius, the larger are its wave number and frequency (energy).  Existence of the beginning point of the spectrum $k=k_{b}$ then  corresponds to a ring of smallest radius $R_{b}$.  The latter can be computed from the $k(R)$ dependence (\ref{vortex_ parametric}).  Indeed, employing extrapolated to zero pressure value of the superfluid density $\rho_{s}=0.145~g/cm^{3}$ \cite{LL9} and the estimate $k_{b}=4.3~\AA^{-1}$ \cite{MP} one finds $R_{b}=3.2~\AA$ \cite{MP}.  
 
High energy part of the vortex branch of the elementary excitation spectrum (\ref{vortex_ parametric}) has been observed in experiments of Rayfield and Reif \cite{RR} who studied the mobility of ions in Helium.  They established that at energies of $1.5$ to $45~eV$ ions create vortex rings, attach to them and move together.  Charged rings of these energies have radii varying from $5\times10^{-6}$ to $10^{-4}~cm$ which correspond to very large wave numbers $k\gg k_{b}$.  So far the beginning part of the vortex branch of the elementary excitations spectrum has not been observed.

The goal of this work is to determine the geometry of wake patterns in unbounded liquid visible in variation of the velocity field due to interference of vortex ring excitations.  Since the motion around the vortex filament is accompanied by a density change of the liquid, there will be an additional density pattern accompanying the velocity field in the wake. As we shall see, as the source velocity $v$ increases, wake patterns undergo qualitative changes which have their origin in the existence of the beginning point of the spectrum $k=k_{b}$ thus opening an opportunity to its observation.

If the requirement $\ln(R/a)\gg 1$ is relaxed to $\ln(R/a)\geqslant 1$ where the expression (\ref{vortex_ parametric}) remains qualitatively correct, one can see that for a ring of radius $R^{*}=ea$ its group $d\omega/dk$ and phase $\omega/k$ velocities coincide.    Corresponding wave number evaluated from the equation for $k(R)$ (\ref{vortex_ parametric}) is $k^{*}=2\pi^{2}\rho_{s}e^{2}a^{2}/m$;  specifically, $d\omega/dk<\omega/k$ if $k>k^{*}$.  Employing experimentally deduced value of $a=0.7~\AA$ \cite{MP,RR} one finds that $k^{*}=1.6~\AA^{-1}$. Since the beginning point of the spectrum is restricted by the inequality $k_{b}>2k_{0}~ (=3.8~\AA^{-1})$ \cite{Pitaevskii66}, we infer that for $k\geqslant k_{b}$ the group velocity is smaller than the phase velocity.  This means that the equation $\omega(k)=kv$ determining boundary values of the wave numbers satisfying the MCL condition (\ref{Cherenkov}) has either one solution or no solutions at all.

With this in mind the effect of the beginning point of the spectrum on wake patterns can be in part anticipated based on the observation that the velocity
\begin{equation}
\label{vortex_velocity}
v_{b}= \frac{\omega_{b}}{k_{b}} 
\end{equation}     
corresponding to the slope of a straight line connecting the origin $\omega(0)=0$ to the beginning point of the vortex branch of the spectrum $\omega(k_{b})\equiv \omega_{b}$ plays a special role.  Indeed, if $v<v_{b}$, then only vortex excitations with sufficiently large wave numbers satisfy the MCL condition (\ref{Cherenkov}) and participate in making wake patterns.  On the other hand, if $v>v_{b}$, the MCL conditions holds for excitations of all allowed wave numbers, $k\geqslant k_{b}$.

A relationship between the Landau critical roton velocity $v_{c}$ (\ref{roton_velocity}) and the velocity $v_{b}$ (\ref{vortex_velocity}) is not yet established since coordinates of the beginning point of the vortex spectrum are not reliably known.  If $v_{b}<v_{c}$, as the source velocity $v$ increases, vortex wake patterns will undergo qualitative change before the rotons of the Landau branch are generated.  The reverse is true if $v_{b}>v_{c}$.  The latter scenario will be realized if the estimates $k_{b}= 4.3~\AA^{-1}$ and $\hbar\omega_{b}=2.2~meV$ \cite{MP} are accurate.  Indeed, extrapolated to zero pressure Landau critical roton velocity (\ref{roton_velocity}) is $v_{c}=60~m/s$ while $v_{b}=79~m/s$.  However even in this case interfering roton and vortex wake patterns can be distinguished thanks to unique features of vortex wake patterns described below.  Regardless of a relationship between $v_{b}$ (\ref{vortex_velocity}) and $v_{c}$ (\ref{roton_velocity}), this opens an experimental opportunity to observe the beginning of the vortex branch of the excitation spectrum.         

Eliminating the ring radius between $\omega$ and $k$, Eqs.(\ref{vortex_ parametric}) can be brought into an explicit $\omega(k)$ dependence:
\begin{equation}
\label{vortex_spectrum}
\omega^{2}= \frac{\pi^{2}\hbar^{2}\rho_{s}}{2m^{3}}k\ln^{2}\left (\frac{mk}{2\pi^{2}\rho_{s}a^{2}}\right ).
\end{equation} 
Since the logarithm is a slowly varying function of its argument and the logarithmic approximation $\ln(R/a)\gg1$ may be only qualitatively correct in the vicinity of the beginning point of the spectrum, the dispersion law of vortex ring excitations will be approximated by an expression
\begin{equation}
\label{spectrum}
\omega^{2}=gk,~~~g\simeq\frac{\hbar^{2}\rho_{s}}{m^{3}},~~~k\geqslant k_{b}
\end{equation}
so that the characteristic velocity (\ref{vortex_velocity}) is given by
\begin{equation}
\label{characteristic_velocity}
v_{b}=\sqrt{\frac{g}{k_{b}}}.
\end{equation}   
If the parameter $g$ would be the free fall acceleration, and existence of the wave number cutoff $k_{b}$ would be ignored, the dispersion law (\ref{spectrum}) would be identical to that of gravity waves on deep water \cite{LL6,Lamb}.  The parameter $g$ in Eq.(\ref{spectrum}) is however eleven orders of magnitude larger than the free fall acceleration which will have dramatic effect on the spatial scale of wake patterns.  

It is straightforward to verify that for the dispersion law (\ref{spectrum})  the MCL condition (\ref{Cherenkov}) becomes $\cos\varphi=\sqrt{g/kv^{2}}$ and can  always be satisfied for sufficiently large wave number $k$.  As a result the vortex wake appears for any velocity, thus implying that $v_{c}=0$. 

The problem of determining wake patterns due to interference of the vortex ring excitations in the approximation given by Eq.(\ref{spectrum}) is then similar to that determining ship wakes.  The important difference lies in the nature of the wave number cutoff.  Indeed, wake patterns due to smooth traveling pressure disturbance can be understood in terms of an effective \textit{ultraviolet} wave number cutoff having its origin in spatial scale (s) of the pressure distribution \cite{CK}.   Similarly, there exists an ultraviolet wave number cutoff in the problem of determining wake patterns due to a charge traversing a two-dimensional electron gas \cite{KS};  here the cutoff originates from the Debye screening.  In both of these cases, the role of the cutoff wavenumber is to suppress sufficiently short-wavelength excitations from participating in forming wake patterns.  

In the case of present interest (\ref{spectrum}), however, the wave number cutoff at $k=k_{b}$ is \textit{infrared} which means sufficiently long-wavelength excitations are excluded from participating in forming wake patterns.  The outcome can be inferred from the classic analysis of wake patterns due to traveling point pressure source \cite{Kelvin,Lamb};  below we follow the approach of Refs. \cite{KCS,CK}.

If $k_{b}=0$ the parameters of the problem such as $g$ and $v$ can be combined into a single length scale 
\begin{equation}
\label{Kelvin_length}
l=\frac{v^{2}}{g}
\end{equation}
called the Kelvin length \cite{CK} which (for point source) determines fine structure of the resulting wake pattern.  For example, for $v\simeq10~m/s$ one finds $l\simeq 1~\AA$ which means that fine structure of the wake can be only resolved with the help of X-ray or neutron scattering. 

In the presence of finite wave number cutoff  $k=k_{b}$ there exists an additional length scale $\simeq k_{b}^{-1}$ which can be combined with the Kelvin length (\ref{Kelvin_length}) to form a dimensionless combination
\begin{equation}
\label{Mach_number}
\mathcal{M}=\sqrt{lk_{b}}=\frac{v}{v_{b}}
\end{equation}
where in the second representation Eqs.(\ref{characteristic_velocity}) and (\ref{Kelvin_length}) were employed.  If $1/k_{b}$ would be a scale characterizing the source, then $\mathcal{M}$ would be a Froude number \cite{Lamb}.  However since the scale $1/k_{b}$ is the property of the medium, it is more appropriate to call the parameter $\mathcal{M}$ a Mach number, the ratio of the source velocity to the characteristic velocity of the medium.  The $\mathcal{M}=0$ case then would be closely related to the original Kelvin wake pattern due to traveling point pressure source \cite{Kelvin,Lamb};  specifically, in any plane intersecting the three-dimensional wake pattern along the path of the source one finds the Kelvin wake \cite{KCS}.  Hereafter the length is measured in units of the Kelvin length (\ref{Kelvin_length}) and wave numbers in units of $1/l$.  Then the constraint $k\geqslant k_{b}$ in Eq.(\ref{spectrum}) becomes $k\geqslant \mathcal{M}^{2}$.

Wake patterns are stationary in the reference frame of the source and their geometry can be determined via Kelvin's stationary phase argument \cite{Kelvin,Lamb,CK,KCS}.  It is a condition of stationary of the phase $f=\textbf{k}\cdot \textbf{r}$ subject to the MCL condition (\ref{Cherenkov}) (here $\textbf{r}$ is the three-dimensional position vector measured relative to the location of the source).  Assuming the source is moving in the positive $x$ direction, the stationary phase condition for the problem at hand (\ref{spectrum}) has the form \cite{CK,KCS}:
\begin{equation}
\label{stationary_phase_general}
-\frac{\varrho}{x}=\frac{\sqrt{k-1}}{2k-1},~~~k\geqslant \mathcal{M}^{2}
\end{equation}
where $\varrho$ is the distance from the $x$-axis (the pattern has axial symmetry around the path of the source).   

Since the phase $f$ is constant along the wavefront, Eq.(\ref{stationary_phase_general}) and $f=\textbf{k}\cdot\textbf{r}$ can be solved relative to $x$ and $\varrho$ to give the equation for the wavefronts in parametric form:
\begin{equation}
\label{parametric}
x(k)=-2\pi n\frac{2k-1}{k^{3/2}},~\varrho(k)=2\pi n\frac{\sqrt{k-1}}{k^{3/2}}, k\geqslant\mathcal{M}^{2}
\end{equation}
where $f=-2\pi n$ and $n$ is positive integer \cite{CK,KCS}.

Eqs. (\ref{stationary_phase_general}) and (\ref{parametric}) can have solutions only for $x<0$ (which is where the wake is) satisfying the inequality $k\geqslant1$ ($k\geqslant 1/l$ (\ref{Kelvin_length}) in the physical units).  The latter fact means that even without the additional constraint $k\geqslant\mathcal{M}^{2}$, wake patterns are a result of interference of sufficiently short-wavelength excitations.    
\begin{figure}
\begin{center}
\includegraphics[width=1\columnwidth]{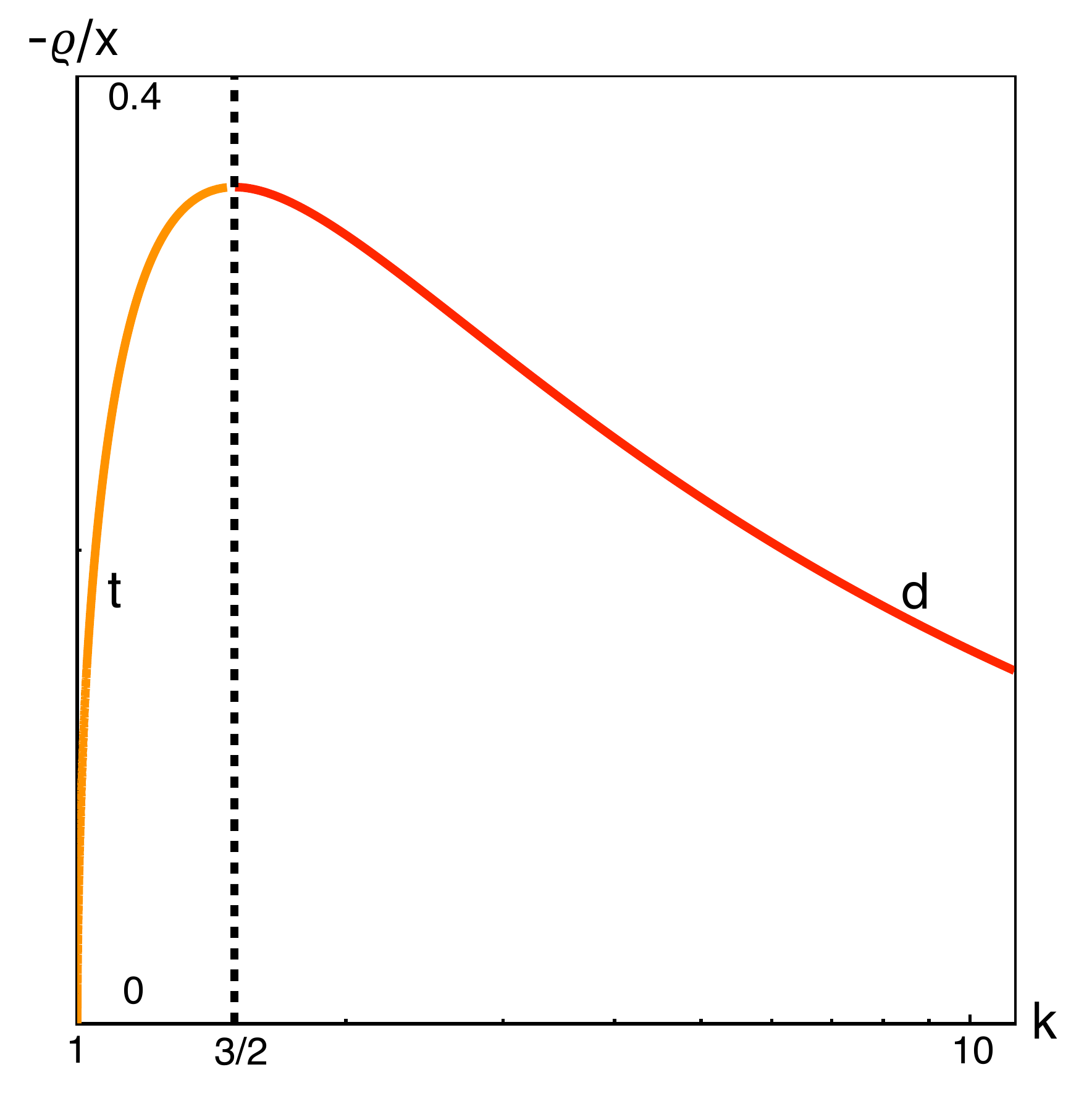}
\caption{(Color online) Right-hand side of the equation of stationary phase (\ref{stationary_phase_general}) (neglecting the $k\geqslant \mathcal{M}^{2}$ constraint) presented in the semi-logarithmic scale as a function of the wave number $k$ (in units of $1/l$ (\ref{Kelvin_length})).  The ascending $1\leqslant k<3/2$ part of the curve (orange) corresponds to transverse (t) wavefronts while its descending counterpart, $k>3/2$, (red) represents diverging (d) wavefronts.} 
\label{stphase}
\end{center}
\end{figure}

The functional dependence (\ref{stationary_phase_general}) is shown in Figure \ref{stphase}.  Its right-hand side vanishes at $k=1$ and $k\rightarrow\infty$ reaching maximum value of $1/2\sqrt{2}$ at $k=3/2$. Thus, Eq.(\ref{stationary_phase_general}) has two solutions for $0\leqslant -\varrho/x < 1/2\sqrt{2}$, corresponding to transverse (t) (ascending orange part of the curve) and diverging (d) (descending red part of the curve) wavefronts. These solutions merge at $-\varrho/x=1/2\sqrt{2}$, while none are found above this value. From this one obtains Kelvin's classic result that the wake is confined by the angle $2\arctan(1/2\sqrt{2}) \approx 39^{\circ}$.  

\begin{figure*}
\begin{center}
\includegraphics[width=0.32\textwidth]{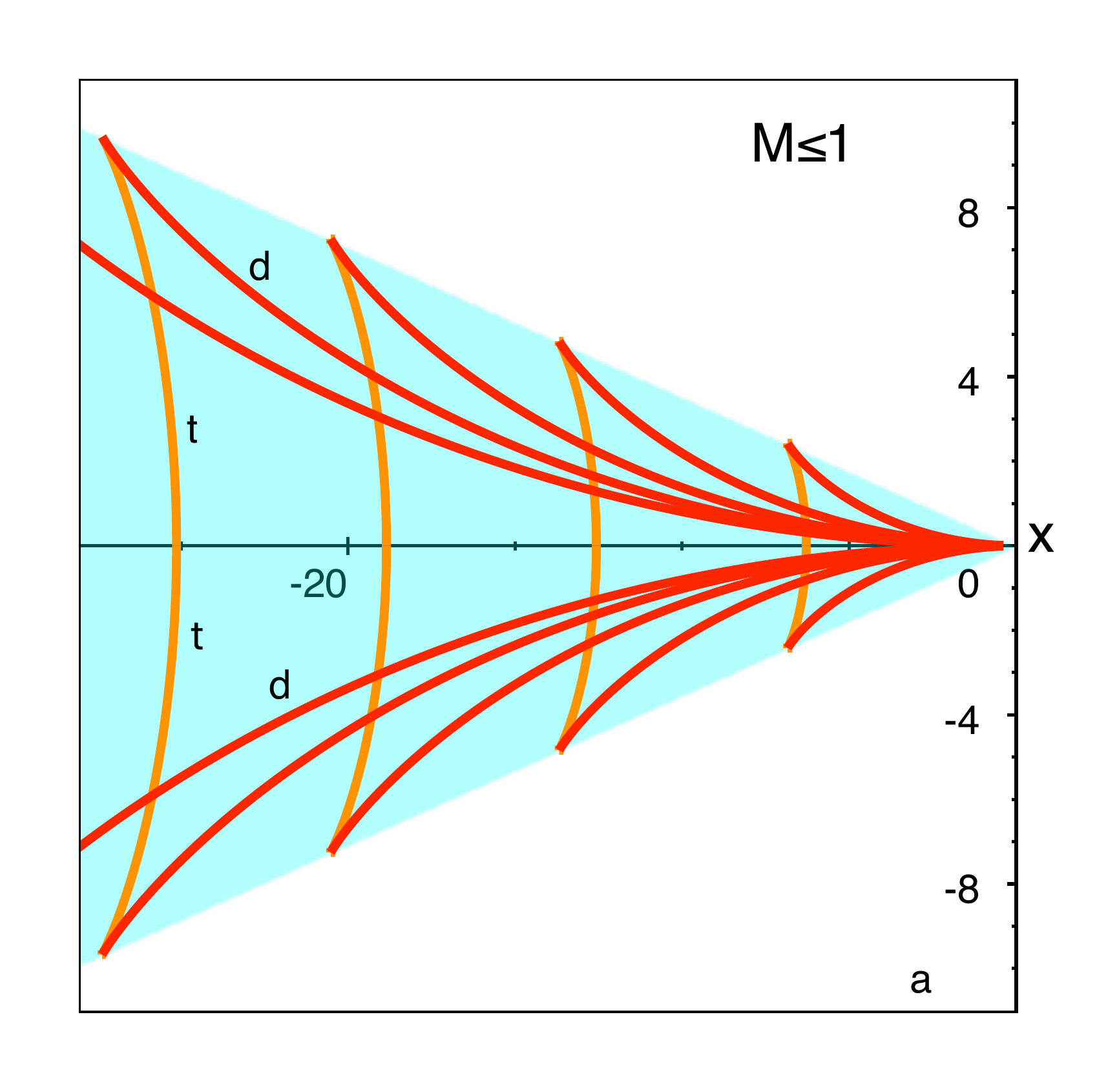}
\includegraphics[width=0.32\textwidth]{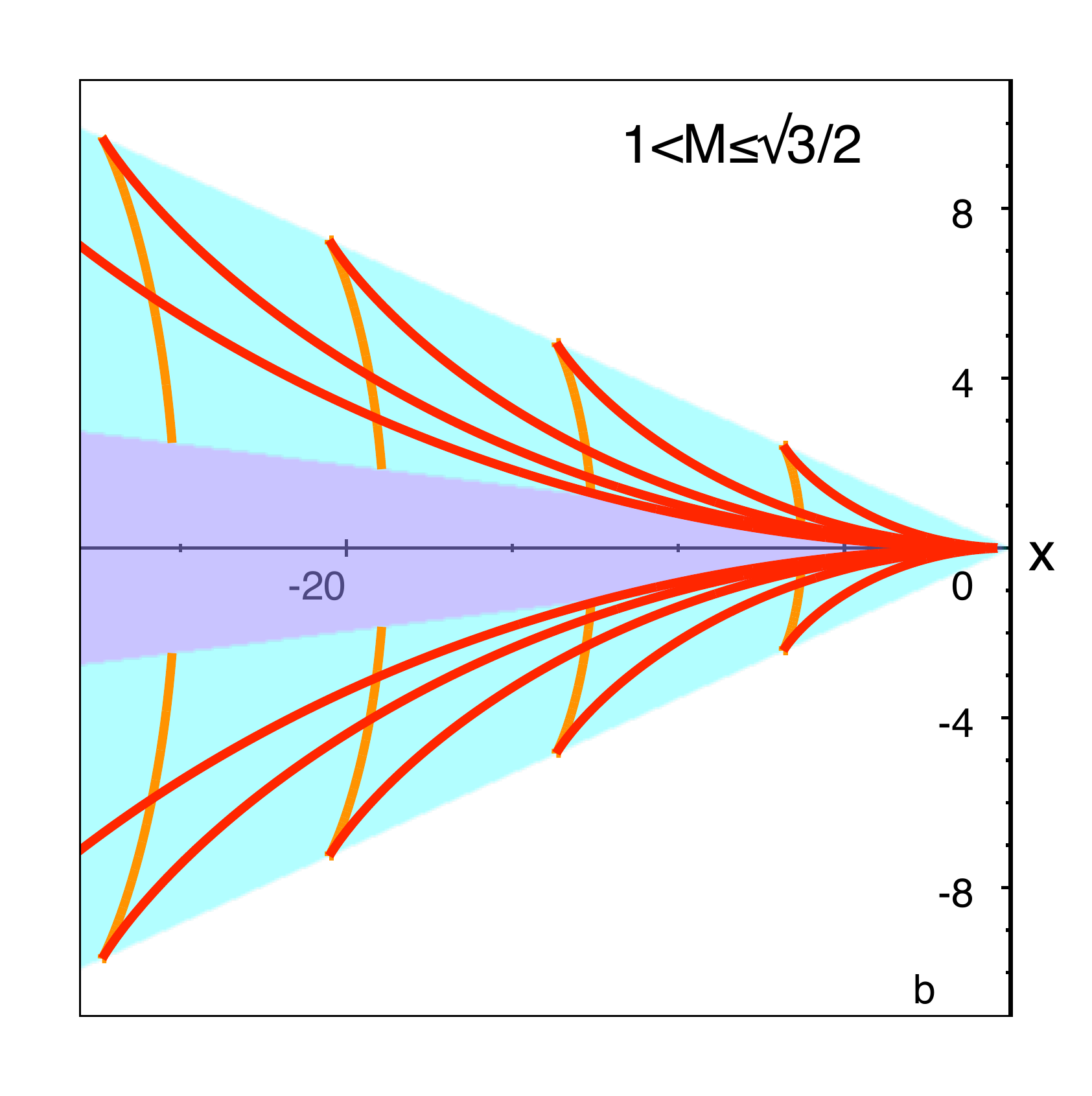}
\includegraphics[width=0.32\textwidth]{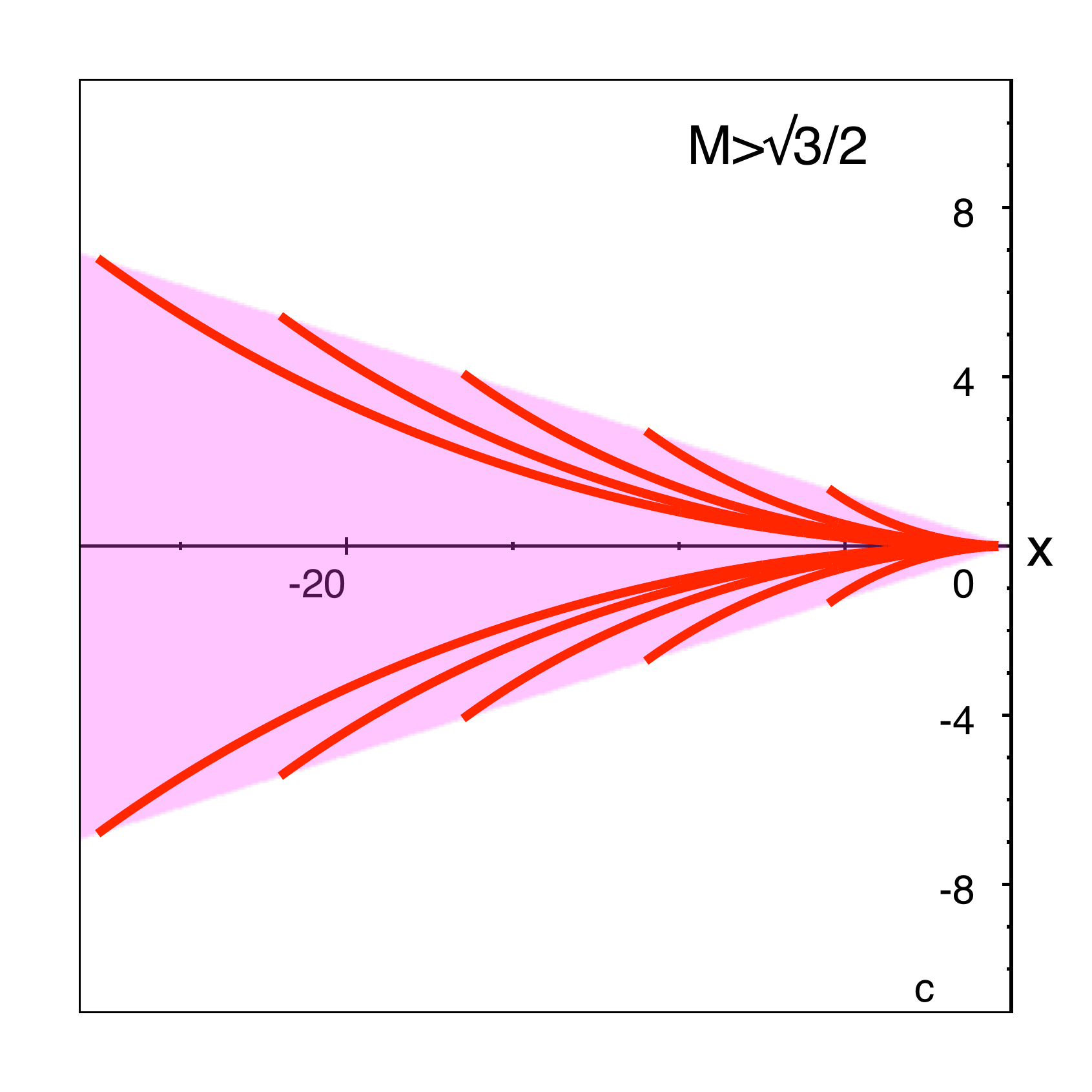}
\caption{(Color online) Evolution of vortex wake patterns in bulk superfluid $^{4}He$ produced by a small source at the origin traveling to the right according to Eqs.(\ref{parametric}) in co-moving reference frame as a function of the Mach  number  $\mathcal{M}$ (\ref{Mach_number}).  Kelvin unit of length (\ref{Kelvin_length}) is utilized.  What is shown is intersection of wave fronts with a plane containing the path of the source.  Transverse (t) and diverging (d) wavefronts are shown in orange and red, respectively.  Regions where both transverse and diverging wavefronts are found are shaded light blue;  regions without transverse wavefronts present are shaded purple (b) or magenta (c).  Opening angle of the conical region that is free of transverse wavefronts, cases (b) and (c), is given by Eq.(\ref{opening_angle}).}
\label{Wakes}
\end{center}
\end{figure*}

The effect of the additional constraint $k\geqslant\mathcal{M}^{2}$ in (\ref{stationary_phase_general}), representing existence of the beginning point of the spectrum (\ref{spectrum}), can be now simply understood.  Indeed, if $\mathcal{M}^{2}<1$ the constraint is unimportant and all the excitations satisfying the inequality $k\geqslant1$ participate in producing the wake which (in any plane containing the path of the source) will be the Kelvin wake.  Specifically, transverse wavefronts connect the edges of the pattern ($k=3/2$), across the path of the source ($k=1$) as shown in Figure 2a in orange.  Their periodicity along the path of the source ($\varrho=0$) in the original physical units, $2\pi l$, is fixed by the Kelvin length (\ref{Kelvin_length}).  Additionally, diverging wavefronts connect the edges of the pattern ($k=3/2$) to the source ($k=\infty$) as shown in Figure 2a in red.  Figures \ref{stphase} and \ref{Wakes} are color coordinated to make it clear that interference of waves whose wave numbers belong to a range marked in given color in Figure \ref{stphase} produces wavefronts colored in the same fashion in Figure \ref{Wakes}.

Wake pattern in Figure \ref{Wakes}a undergoes qualitative change when the Mach number $\mathcal{M}$ exceeds unity because now the constraint $k\geqslant \mathcal{M}^{2}$ is stronger than the built-in condition $k\geqslant1$ in Eqs.(\ref{stationary_phase_general}) and (\ref{parametric}).  The critical value $\mathcal{M}=\mathcal{M}_{1}=1$ corresponds to the source velocity $v=v_{1}$ matching the characteristic velocity (\ref{vortex_velocity}) as was already anticipated.  If $\mathcal{M}>1$, there are still two cases to consider:

(i)  If $1<\mathcal{M}\leqslant \sqrt{3/2}$, then transverse wavefronts no longer reach the path of the source.  According to the condition of stationary phase (\ref{stationary_phase_general}) they start out on the conical surface 
\begin{equation}
\label{intermediate}
-\frac{\varrho}{x}=\frac{\sqrt{\mathcal{M}^{2}-1}}{2\mathcal{M}^{2}-1}
\end{equation}  
corresponding to $k=\mathcal{M}^{2}$ and extend to the edges of the pattern, $k=3/2$.  Transverse wavefronts are absent within the cone (\ref{intermediate}) whose opening angle is given by   
\begin{equation}
\label{opening_angle}
2\theta=2\arctan\frac{\sqrt{\mathcal{M}^{2}-1}}{2\mathcal{M}^{2}-1}
\end{equation}
The outcome is shown in Figure \ref{Wakes}b.  When experimentally observed, change in appearance of wake patterns between Figures \ref{Wakes}a and \ref{Wakes}b would be an indirect evidence of the beginning part of the vortex branch of the elementary excitations spectrum. 

(ii)  Wake pattern in Figure \ref{Wakes}b undergoes another qualitative change when the Mach number $\mathcal{M}$ exceeds $\sqrt{3/2}$ because now transverse wavefronts completely disappear.  The critical value $\mathcal{M}=\mathcal{M}_{2}=\sqrt{3/2}$ corresponds to the source velocity $v=v_{2}=\sqrt{3/2}v_{b}\approx 97~m/s$ where we used already mentioned estimate $v_{b}=79~m/s$.  For $\mathcal{M}>\sqrt{3/2}$ the wake is made only of diverging wavefronts as shown in Figure \ref{Wakes}c.  The entire wake pattern is now confined within a cone defined by Eq.(\ref{intermediate}) whose opening angle (\ref{opening_angle}) decreases with the Mach number;  in the $\mathcal{M}\gg1$ limit the opening angle vanishes as $1/\mathcal{M}$.

The instants $v=v_{1}$ and $v=v_{2}$ when vortex wake patterns undergo qualitative changes in their appearance represent critical phenomena.  It is expected that they will be accompanied by singular changes in the wave resistance which we are planning to study in the future.   

The property of superfluidity is to some extent eliminated within the region occupied by vortex wake patterns.  We argue that superfluidity is completely eliminated at the wavefronts shown in Figure \ref{Wakes} which is where the vorticity is largest.

To summarize, we demonstrated that vortex wake patterns in superfluid $^{4}He$ exhibit elements of Kelvin ship wake to varying degree depending on the source speed.  We explained that changes in appearance of wake patterns with the source speed can serve as an experimental evidence of the beginning part of the vortex excitation spectrum.  It is necessary to emphasize that coarse-grained outlines of the wake patterns in Figures \ref{Wakes}a-\ref{Wakes}c should be observable by light scattering techniques while the fine structure of all the discussed wake patterns can be only resolved with the help of $X$-ray or neutron scattering.

These results (with suitable changes) also apply to other types of superfluids, specifically fermion superfluids in cold atomic gases \cite{Rishi} and dipolar quantum gases \cite{dip} thus enlarging the pool of experimental systems where vortex wake patterns can be observed.

Finally, the idea that fingerprints of relevant elementary excitations can be found in experimentally observed wake patterns goes beyond the example of vortex rings in superfluid $^{4}He$ investigated in this work.  Other notable examples include wake patterns in superfluid $^{4}He$ due to the Landau phonon-roton excitations \cite{KCS} and Kelvin-Mach wakes in a two-dimensional Fermi gas;  the latter have their origin in interference of plasma oscillations \cite{KS}.  Mathematically accurate solution of the problem of determination of the spectrum of elementary excitations from wake patterns may have an applied value.     
     
The author is grateful to J. P. Straley for valuable comments.

\end{document}